# Decoupling the effects of composition and strain on the vibrational modes of GeSn

É. Bouthillier,[1] S. Assali,[1] J. Nicolas,[1] and O. Moutanabbir[1]

[1] Department of Engineering Physics, École Polytechnique de Montréal, C. P. 6079, Succ. Centre-Ville, Montréal, Québec H3C 3A7, Canada

**ABSTRACT:**

We report on the behavior of Ge-Ge, Ge-Sn, Sn-Sn like and disorder-activated vibrational modes in GeSn semiconductors investigated using Raman scattering spectroscopy. By using an excitation wavelength close to $E_1$ gap, all modes are clearly resolved and their evolution as a function of strain and Sn content is established. In order to decouple the individual contribution of content and strain, the analysis was conducted on series of pseudomorphic and relaxed epitaxial layers with a Sn content in the 5-17at.% range. All vibrational modes display qualitatively the same behavior as a function of content and strain, *viz.* a linear downshift as the Sn content increases or the compressive strain relaxes. Simultaneously, Ge-Sn and Ge-Ge peaks broaden, and the latter becomes increasingly asymmetric. This asymmetry, coupled with the peak position, is exploited to implement an empirical approach to accurately quantify the Sn composition and lattice strain from Raman spectra.

Understanding the behavior of different vibrational modes in a semiconductor is of paramount importance to probe its crystal phase and symmetry, composition, lattice strain, isotopic content, electronic and phononic properties.[1–3] In this regard, Raman scattering spectroscopy has thus become an ubiquitous characterisation technique as information-rich spectra are acquired from straightforward and non-destructive measurements. Therefore, it is commonly used to evaluate the chemical composition and lattice properties of, for instance, group-IV semiconductors such as strained Si,[4–6] strained Ge,[7–10] SiGe,[11–14] and GeSn layers.[15–25] The latter are particularly of growing interest because of their relevance to Si-compatible light emission and detection applications in the short- and mid-wavelength infrared,[26–35] which can lead to the integration of optoelectronic and photonic circuits on complementary metal-oxide-semiconductor (CMOS) platforms.[36–38]

Previous reports on the vibrational modes of GeSn mainly focused on Ge-Ge longitudinal optical (LO) mode as the analyses relied on the use of 488 nm[15,16] or 532 nm[18–24] excitation lines. Under these conditions, the signal-to-noise ratio is too low to clearly distinguish Sn-related vibrational modes in the vicinity of the more prominent Ge-Ge LO peak. This also applies to the study of ternary SiGeSn semiconductors.[39–41] When using a 633 nm excitation laser, the signal-to-noise ratio is significantly enhanced, thus allowing a clear distinction of Ge-Ge and Ge-Sn modes, in addition to other features such as disorder-activated (DA) and Sn-Sn like modes. This higher sensitivity is attributed to the increase in Raman scattering cross section when the excitation wavelength becomes close to the material's $E_1$ gap.[39,42] Oehme et al.[25] and D'Costa et al.[17] provided a quantitative description of the evolution of peak positions as a function of the composition. However, in these studies, the investigated samples are either pseudomorphic[25] or relaxed.[17] Consequently, strain and compositional effects cannot be fully decoupled. In this work, we address this very issue by investigating the individual effects of strain and composition on the behavior of Ge-Ge, Ge-Sn, DA and Sn-Sn like vibrational modes in GeSn alloys using both pseudomorphic and relaxed layers with Sn content in the 5-17at.% range. This ability to distinguish the individual contribution of each parameter allows us to develop an approach to independently evaluate strain and composition of GeSn layers from Raman measurements by exploiting the properties of vibrational modes and lattice disorder.

The investigated GeSn samples were grown on 0.6-1.1µm-thick Ge virtual substrates (VS) on 4-inch Si (100) wafers in a low-pressure chemical vapor deposition (CVD) reactor, with ultra-pure $H_2$ as carrier gas and 10% monogermane ($GeH_4$) and tin-tetrachloride ($SnCl_4$) precursors.[43,44] The Sn content ($y$), the residual in-plane strain ($\varepsilon$), and the degree of strain relaxation ($R$) for all layers, listed in Table I, were estimated from high-resolution X-Ray Diffraction (XRD) Reciprocal Space Mapping (RSM) measurements, applying a bowing parameter of 0.041Å.[44,45] Micro-Raman scattering spectroscopy was performed using an InVia Raman Microscope from Renishaw with a 633nm laser and an 1800/mm grating. In other studies, the use of shorter wavelength laser was sometimes justified by the need for a shallower penetration depth to avoid the background signal from underlying Ge or GeSn buffer layers.[39] This is not a concern here even when a 633nm excitation laser is used. Indeed, the comparison of the penetration depths and thicknesses estimated from ellipsometry and transmission electron microscopy (TEM) confirmed that the investigated layers are sufficiently thick to supress the contribution from the underlying layers to the measured signal. While Voigt or Lorentzian[18] functions are commonly used to fit Raman peaks, they cannot reproduce the asymmetric behavior that is typical to Raman modes of semiconductor alloys.[46,47] This asymmetry is due to alloying as the substitution of Sn atoms in the lattice breaks the translational symmetry and leads to a relaxation of the $\vec{q} = \vec{0}$ momentum selection rule.[48] To better reproduce the line shape of the LO modes, we employed exponentially modified gaussian (EMG) functions.[17,49] Note that the extracted peak positions correspond to the wavenumbers of maximal intensity.

Two sets of pseudomorphic and relaxed GeSn layers were investigated in this study, in addition to layers with intermediate strain relaxation. For the pseudomorphic GeSn series, a 4.0at.% bottom layer (BL) was first grown at 350°C, then the temperature was decreased to grow the top layer (TL) at a higher Sn content in the 9-13at.% range (samples A-D). The other parameters were kept constant during growth. Fig. 1a exhibits a typical scanning TEM (STEM) image for sample D with a 50nm-thick TL (13at.%) on a 45nm-thick BL (4at.%). As the critical thickness for strain relaxation increases with decreasing Sn content[50] the interaction of the laser beam only takes place in the TL in all pseudomorphic samples. The corresponding RSM map in Fig. 1b shows that the GeSn BL and TL are pseudomorphic with respect to Ge-VS, corresponding to a $R < 5\%$. Similar $R$ was estimated for all samples regardless of the growth temperature. The presence of interference fringes in Fig. 1b, also observed in the $2\theta - \omega$ scan around the (004) XRD order, indicates a high degree of crystallinity. Note that the thickness of the TL is in the 40-50nm range.

For the relaxed series, 500-700nm-thick GeSn layers were grown at a fixed temperature in the 330-300°C range, leading to a composition in the 7-13at.% range (samples E-H). In the TEM image for a 13at.% Sn layer (Fig. 1c) dislocations are mainly observed within the first 200-300 nm of the GeSn layer.[44] The associated RSM map shows strong broadening along $q_x$ resulting from plastic relaxation (Fig. 1d),[44] with an estimated $R$ exceeding 75% in all samples (Fig. 1e). The reduced compositional grading at thicknesses above 300 nm leads to a uniform strain and composition profile within the depth probed in Raman measurements (~30nm). The Sn content for the pseudomorphic and relaxed sets are plotted in Fig. 1f as a function of the growth temperature. An increase in Sn content of 1.3±0.3at.% for every 10°C decrease in growth temperature is observed for the pseudomorphic samples, while a higher rate of –(2.1±0.2at.%)/10°C is estimated for the relaxed layers, resulting from enhanced strain minimization during growth.[44,51] In addition to pseudomorphic and relaxed sets of samples described above, Raman analysis was also extended to six other samples of various values of strain and composition (Table I).

Fig. 2a-b displays representative polarized Raman spectra, recorded for sample F ($y = 8.8$at.%, $\varepsilon = -0.25\%$). For comparison, polarized Raman spectra of a Ge-VS are also displayed in Fig. 2c. For both samples, the main Ge-Ge mode (~300cm$^{-1}$) is much stronger under $x(z'z')\bar{x}$ and $x(zy)\bar{x}$ configurations as compared to $x(zz)\bar{x}$ and $x(z'y')\bar{x}$ configurations, which is consistent with the selection rules of LO phonons. For GeSn layer, the Ge-Sn LO mode[17,25] (dotted vertical line) shows the same behavior. The shoulder visible on the low wavenumber side of the Ge-Ge peak (dashed vertical line) is related to DA mode attributed to a maximum in the one-phonon density of states in Ge.[17,52,53] Its presence becomes more apparent under $x(zz)\bar{x}$ configuration because of the low intensity of the two adjacent LO modes. This additional contribution is accounted for in the fit of Raman spectra, which provides an accurate identification of the characteristics of each mode. An example of a fitted unpolarized spectrum in the 235-325cm$^{-1}$ range is displayed in Fig. 2d. In addition, second-order Ge-Ge modes are also detected in $x(z'z')\bar{x}$ and $x(zz)\bar{x}$ configurations for Ge-VS (2TA between 100cm$^{-1}$ and 250cm$^{-1}$ [25,54], TO+TA near 350cm$^{-1}$ and 2LA near 380cm$^{-1}$ [55]). These modes also appear with the same selection rules in the GeSn sample, as well as an additional mode near 180cm$^{-1}$ (dash-dotted vertical line). Since this peak doesn't follow the selection rules of a LO mode, it cannot be straightforwardly assigned to the Sn-Sn mode expected near this frequency.[56,57] We believe the small intensity of Sn-Sn vibrations, if present, is hindered by another mode, which we attribute to a DA peak, because its frequency coincides with a maximum in the phonon density of states of Ge[17,53,58] and its intensity for the different polarizations changes in a similar fashion as for the DA peak near 280cm$^{-1}$. For this reason, we refer to this mode as Sn-Sn like, in concordance with Refs.[17,25]

Raman spectra of the pseudomorphic and relaxed GeSn samples with variable Sn compositions are displayed in Fig. 3. Interestingly, as Sn content increases, the peak positions in the pseudomorphic layers remain almost unaffected by the change in the Sn content (Fig. 3a), whereas a progressive shift to lower wavenumbers is observed for both modes in the relaxed layers (Fig. 3b). Due to the relatively larger atomic mass of Sn, when a significant amount of Sn atoms is incorporated into Ge lattice, a downshift of Ge-Ge and Ge-Sn modes is expected. This behavior is visible in the relaxed layers, while in the pseudomorphic layers this anticipated downshift is counterbalanced by the upshift associated with the increased compressive strain due to the higher Sn content in the lattice.

Examples of three-dimensional plots of the Raman shifts $\omega$ (black spheres) for Ge-Ge and Ge-Sn modes as a function of Sn content and strain are shown in Fig. 4a-b. The measured data points belong to the same planes, which can be described by two-dimensional linear regressions

$$\omega = \omega_0 + ay + b\varepsilon, \qquad (1)$$

where $\omega_0$ is a characteristic wavenumber, and $a$ and $b$ are the fitting parameters. The resulting fits, superimposed on the scatter plots, and the projections on the 2D space in Fig. 4c-d confirm that the linear regressions accurately represent the behavior of the four peaks in all studied samples.

The fitting coefficients are listed in Table II and the errors are calculated considering 95% confidence intervals. The coefficients of determination being larger than 0.94 and the relatively small errors on both $a$ and $b$ slopes indicate that the planar fits describe adequately the mode distribution. Furthermore, the calculated $\omega_{0,GeGe}$ is equal to the value obtained for bulk Ge and the $a$ and $b$ slopes are comparable to those found in earlier studies of Ge-Ge mode.[18,47] As for the Ge-Sn mode, $a$ and $b$ values are slightly higher than those of Ge-Ge mode. However, the slopes of all four peaks are remarkably close and sometimes overlap when considering the uncertainties.

The analysis above demonstrates that the peak positions of the Ge-Ge, Ge-Sn, DA and Sn-Sn like modes evolve qualitatively similarly. The behavior of the integrated intensity, width, and asymmetry of each one of these vibrational modes was also evaluated. We found that an increase in Sn content is associated with an increase in the relative integrated intensity of Ge-Sn and DA modes. The latter also increases as the layers become more compressively strained. Furthermore, we also noticed that the full width at the half maximum (FWHM) of both Ge-Ge and Ge-Sn peaks increase with higher Sn content and higher relaxation, as expected from the increase in lattice disorder. As for the asymmetry parameter $t$ of the Ge-Ge peak, included in Table II, it increases with Sn content with a relatively strong correlation ($R^2 = 0.9533$). The strain, however, has no measurable impact on $t$. This explains why the DA peak is less prominent for samples with high Sn contents. The increase of the Ge-Ge peak asymmetry, together with a shift of the peak position to lower wavenumbers, results in a larger superimposition of the two contributions. Nevertheless, this superimposition remains partial and it is still possible to discriminate them, as confirmed by the high $R^2$ obtained in the two-dimensional linear regressions.

Any pair of equations describing peak positions as a function of $y$ and $\varepsilon$ would technically be sufficient to estimate the composition and strain of a GeSn layer based on its Raman spectrum. However, since the positions of all modes evolve remarkably similarly, this becomes challenging based solely on peak positions. In fact, due to the comparable $a$ and $b$ slopes, the strain and composition estimation obtained when solving equations (1) for a pair of peaks will be highly dependent on small changes in the input parameters $\omega$. However, as previously stated, the composition and strain do not only affect the peak positions, but also the peak areas, widths, and asymmetry. For instance, the two-dimensional linear regression of $t$ for the Ge-Ge peak results in $a$ and $b$ coefficients that are very different from those obtained for the peak positions (Table II). Therefore, the composition and strain of GeSn alloys can be extracted directly by solving a set of two equations describing the behavior of the peak position and asymmetry of the main Ge-Ge mode,

$$\omega_{Ge-Ge} = \omega_{0,Ge-Ge} + a^{\omega}_{Ge-Ge}y + b^{\omega}_{Ge-Ge}\varepsilon, \qquad (2)$$

$$t_{Ge-Ge} = t_{0,Ge-Ge} + a^{t}_{Ge-Ge}y + b^{t}_{Ge-Ge}\varepsilon, \qquad (3)$$

This approach allowed to retrieve values which are close to those measured by XRD, as shown in Table I. This is a clear demonstration that combining peak position and asymmetry is sufficient for an accurate analysis of strain and composition in GeSn semiconductors using Raman spectroscopy. To test this approach, we carried out Raman line-scans on under-etched GeSn microdisks (not shown). The model enabled the evaluation of the bulk composition with an accuracy of 5at.% in addition to providing the local residual strain across the microdisks, thus clearly demonstrating the relevance of micro Raman spectroscopy to probe the local composition and strain in individual GeSn micro- and nano-structures.

In summary, we described a detailed investigation of GeSn vibrational modes at variable strain and content. The clear distinction of Raman features was made possible by the use of a 633nm laser. The broad range of Sn content of CVD-grown GeSn layers at different relaxation levels allowed the decoupling of the individual effects of composition and strain on the behavior of Raman vibrational modes. We found that Ge-Ge, Ge-Sn, DA and Sn-Sn like modes all downshift as Sn content increases. Strain relaxation induces qualitatively the same behavior. We also found that the peak positions alone are not sufficient to estimate the content and strain from Raman spectra. This limitation was circumvented by exploiting the alloying-induced asymmetry of the Ge-Ge peak as an additional parameter to simultaneously obtain the strain and Sn content in GeSn layers. These results lay the groundwork to employ Raman spectroscopy for a non-destructive characterization of GeSn-based structures and devices.

**ACKNOWLEDGMENTS**

The authors thank S. Mukherjee and A. Attiaoui for the fruitful discussions, and J. Bouchard for the technical support with the CVD system. O.M. acknowledges support from NSERC Canada (Discovery, SPG, and CRD Grants), Canada Research Chair, Canada Foundation for Innovation, Mitacs, and PRIMA Québec.

**Notes**

The authors declare no competing financial interest.

**TABLES CAPTIONS**

**Table I**. List of samples investigated in this work. The growth temperature $T$, the composition and strain values as estimated from XRD and Raman measurements are shown, as well as the wavenumbers of the Sn-Sn like, Ge-Sn, DA, and Ge-Ge modes.

**Table II.** Results of the two-dimensional linear regressions performed for the position of each mode and for the asymmetry parameter $t$ of the Ge-Ge peak.

**FIGURES CAPTIONS**

**Figure 1.** (a-b) STEM image (a) and RSM around the asymmetrical (224) reflection (b) for the pseudomorphic sample D. (c-d) TEM image (c) and RSM map (d) for the relaxed sample H. (e-f) Relaxation ratio $R$ (e) and Sn content (f) as a function of the growth temperature.

**Figure 2.** (a-b-c) Raman signal of GeSn (sample F) (a-b) and Ge-VS (c) for different polarizations; (d) Fit of the Ge-Ge, DA, and Ge-Sn modes for sample F. The coefficient of determination $R^2$ is between 0.9962 and 0.9994 for all samples. Note: $\hat{x}//[100], \hat{y}//[010], \hat{z}//[001], \hat{y}'//[011], \hat{z}'//[0\bar{1}1]$.

**Figure 3.** Ge-Sn and Ge-Ge Raman modes recorded for pseudomorphic (a) and relaxed (b) series.

**Figure 4.** (a-b) Planar regressions for the position of Ge-Ge (a) and Ge-Sn (b) modes. (c-d) Projections of the planar regressions of the four modes in a 2D space. The vertical axis corresponds

to the Raman shift corrected with the effect of strain (c) or composition (d). The lines correspond to side views of the planes. Errors bars are smaller than symbols.

**Table I**

| Sample | | T (°C) | $y_{XRD}$ (at.%) | $y_{Raman}$ (at.%) | $\varepsilon_{XRD}$ (%) | $\varepsilon_{Raman}$ (%) | $\omega_{Sn-Sn}$ (cm$^{-1}$) | $\omega_{Ge-Sn}$ (cm$^{-1}$) | $\omega_{DA}$ (cm$^{-1}$) | $\omega_{Ge-Ge}$ (cm$^{-1}$) |
|---|---|---|---|---|---|---|---|---|---|---|
| Pseudo-morphic | A | 310 | 8.8 | 9.3 | -1.17 | -1.30 | 182.9 | 260.2 | 290.72 | 298.99 |
| | B | 300 | 10.6 | 10.1 | -1.39 | -1.40 | 181.5 | 259.4 | 291.16 | 298.83 |
| | C | 290 | 12.0 | 11.8 | -1.60 | -1.64 | 180.8 | 258.8 | 291.36 | 298.60 |
| | D | 280 | 13.0 | 13.3 | -1.79 | -1.84 | 179.8 | 258.6 | 291.71 | 298.31 |
| Relaxed | E | 330 | 6.8 | 7.1 | -0.14 | -0.19 | 181.9 | 256.6 | 288.86 | 295.38 |
| | F | 320 | 8.8 | 8.3 | -0.25 | -0.13 | 179.7 | 255. | 288.04 | 294.11 |
| | G | 310 | 10.9 | 11.4 | -0.30 | -0.41 | 178.1 | 254.78 | 287.02 | 292.89 |
| | H | 300 | 13.2 | 13.2 | -0.39 | -0.51 | 176.3 | 254.52 | 286.63 | 291.87 |
| Other | I | 340 | 4.6 | 4.1 | -0.23 | -0.03 | 183.2 | 259.19 | 289.37 | 297.18 |
| | J | 300 | 13.3 | 13.2 | -0.33 | -0.36 | 175.1 | 254.29 | 286.11 | 291.13 |
| | K | 280 | 15.4 | 16.1 | -1.17 | -1.32 | 175.6 | 254.92 | 287.85 | 293.43 |
| | L | 280 | 16.0 | 16.1 | -1.41 | -1.25 | 175.2 | 254.76 | 287.82 | 293.06 |
| | M | 280 | 16.1 | 17.2 | -1.16 | -1.38 | 175.0 | 254.50 | 287.44 | 292.80 |
| | N | 280 | 16.9 | 14.9 | -1.27 | -0.85 | 173.7 | 254.09 | 287.39 | 292.14 |

**Table II**

| Regression | $\omega_0$ or $t_0$ (cm$^{-1}$) | $a$ (cm$^{-1}$) | $b$ (cm$^{-1}$) | $R^2$ |
|---|---|---|---|---|
| $\omega_{Ge-Ge}$ | 300.4 ± 0.9 | -84 ± 8 | -491 ± 52 | 0.9820 |
| $\omega_{DA}$ | 291.3 ± 0.7 | -49 ± 7 | -347 ± 45 | 0.9685 |
| $\omega_{Ge-Sn}$ | 261 ± 1 | -68 ± 11 | -347 ± 69 | 0.9485 |
| $\omega_{Sn-Sn}$ | 188 ± 1 | -104 ± 9 | -292 ± 59 | 0.9822 |
| $t_{Ge-Ge}$ | 1.0 ± 0.1 | 6 ± 1 | -5 ± 7 | 0.9533 |

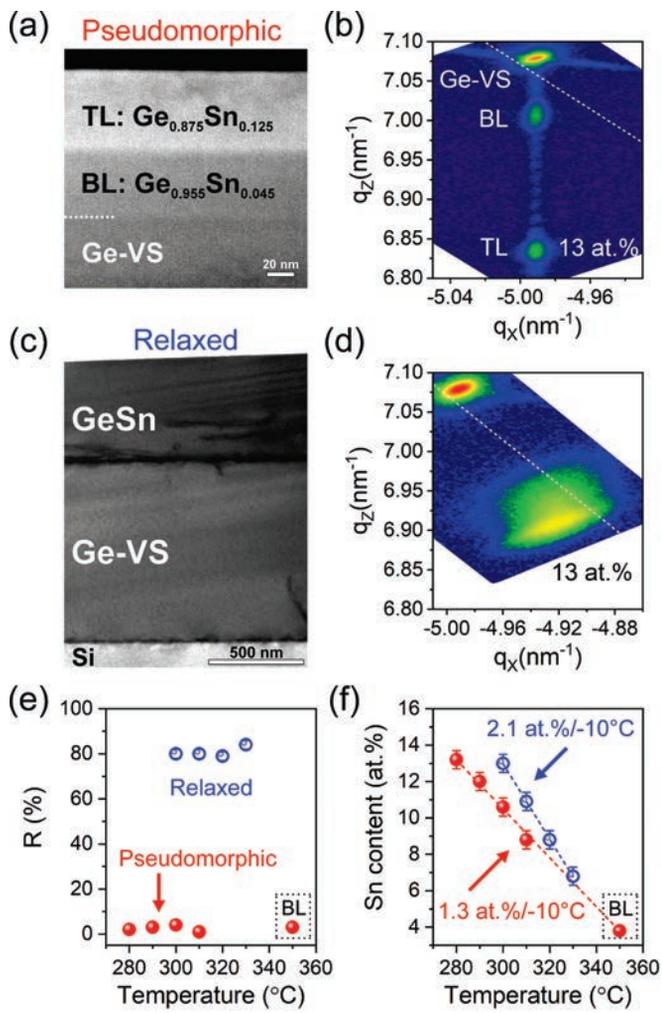

**Figure 1**

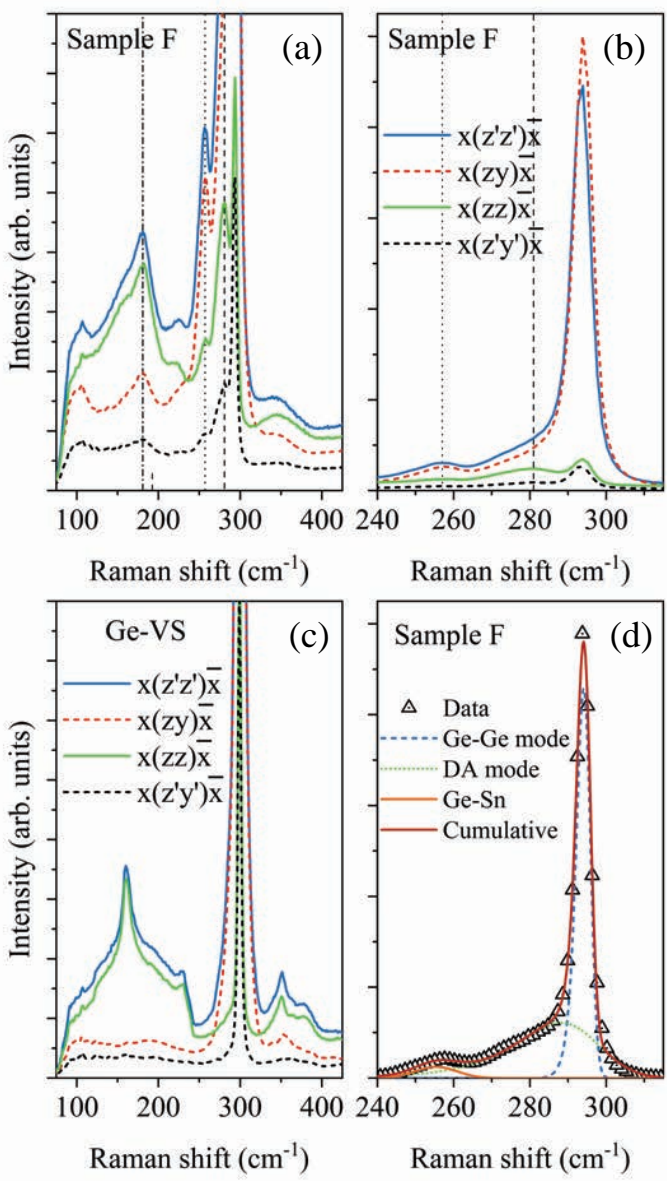

**Figure 2**

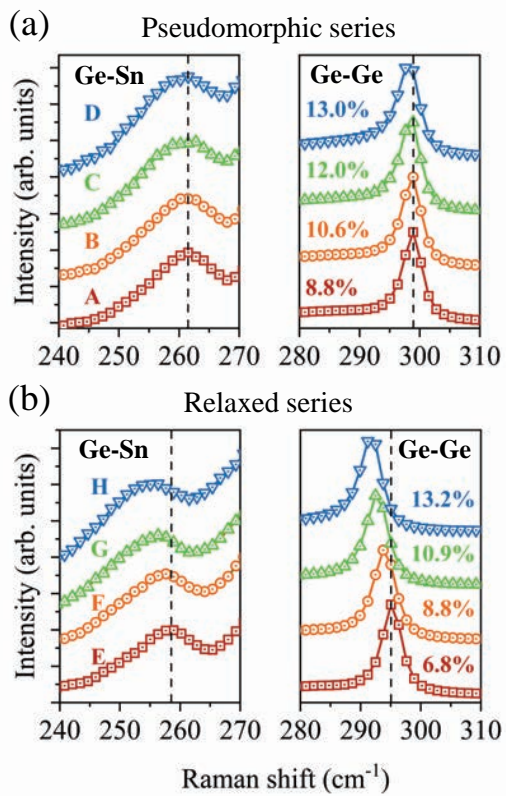

**Figure 3**

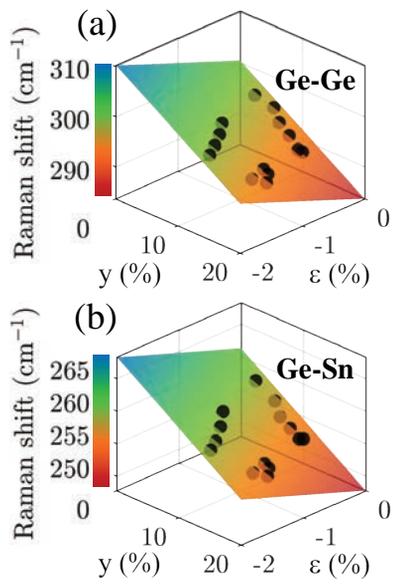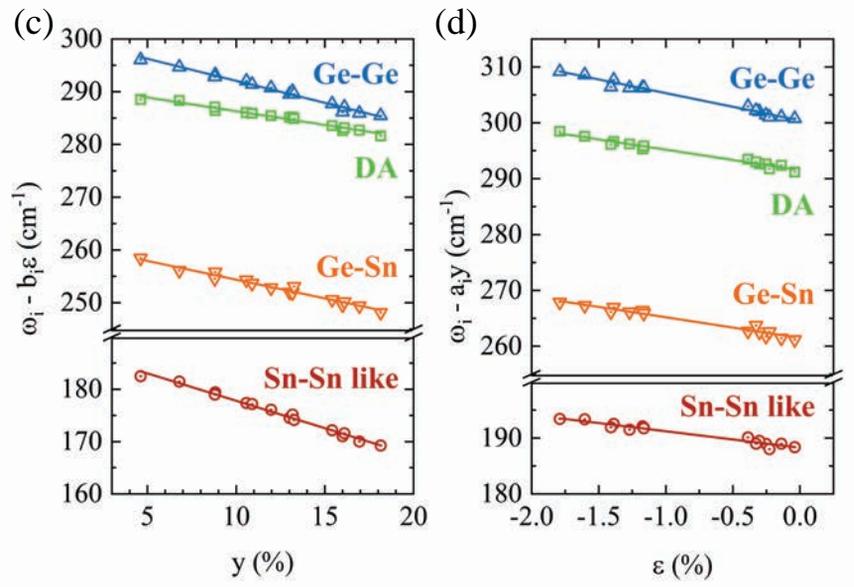

Figure 4